\documentclass[11pt]{article}
\usepackage{amsfonts}
\usepackage{amsmath}
\usepackage{geometry}
\usepackage[colorlinks=true, linkcolor=red, citecolor=blue]{hyperref}
\usepackage{float}
\usepackage{mathtools}
\usepackage{caption}
\usepackage{graphicx}
\usepackage[font=small,labelfont=bf,labelsep=space]{caption}
\usepackage{subfig}
\usepackage{setspace}

\makeatletter

\@addtoreset{equation}{section}
\makeatother

\input{tcilatex}
\begin{document}

\title{Polyanalytic Hermite polynomials associated \\
with the elliptic Ginibre model}
\author{Nizar Demni$^{\ast }$ and Zouha\"{\i}r Mouayn$^{\diamond ,\natural }$
\\
%EndAName
$^{\ast }${\footnotesize \ Aix-Marseille University CNRS Centrale Marseille
I2M-UMR 7373, }\\
{\footnotesize 39 Rue F. Joliot Curie, 13453 Marseille, France\vspace*{-0.2em%
}}\\
{\footnotesize \ e-mail: nizar.demni@univ-amu.fr}\\
$^{\diamond }${\footnotesize \ Department of Mathematics, Faculty of
Sciences and Technics (M'Ghila),\vspace*{-0.2em}}\\
{\footnotesize \ Sultan Moulay Slimane University, BP. 523, B\'{e}ni Mellal
23000, Morocco }\\
$^{\natural }${\footnotesize Institut des Hautes \'{E}tudes Scientifiques,
Paris-Saclay University,}\\
{\footnotesize 35 route de Chartres, 91893 Bures-sur-Yvette, France.}\\
{\footnotesize e-mail: mouayn@usms.ma, mouayn@ihes.fr}}
\maketitle

\begin{abstract}
Motivated by the connection between the eigenvalues of the complex Ginibre
matrix model and the magnetic Laplacian in the complex plane, we derive
analogues of the complex Hermite polynomials for the elliptic Ginibre model.
To this end, we appeal to squeezed creation and annihilation operators
arising from the Bogoliubov transformation of creation and annihilation
operators on the Bargmann-Fock space. The obtained polynomials are then
expressed as linear combinations of products of Hermite polynomials and
share the same orthogonality relation with holomorphic Hermite polynomials.
Moreover, this expression allows to identify them with the $2D$-Hermite
polynomials associated to a unimodular complex symmetric $2\times 2$ matrix.
Afterwards, we derive, for any Landau level, a closed formula for the kernel
of the isometry mapping the basis of (rescaled) holomorphic Hermite
polynomials to the corresponding complex Hermite polynomials. This kernel is
also interpreted in terms of the two-photon coherent states and the
metaplectic representation of the $SU(1,1)$ group.
\end{abstract}

\section{Introduction and main results}

\subsection{The Ginibre model and complex Hermite polynomials}

The complex Hermite polynomials $(H_{m,n}(z,\overline{z}))_{m,n\geq 0}$,
also called It\^{o} Hermite polynomials, arise in various contexts including
probability theory, combinatorics and mathematical physics (see e.g. the
review paper \cite{Ism-Zha}). They are planar polynomials in $(z,\overline{z}%
)$ which are orthogonal with respect to the two-dimensional Gaussian measure 
\begin{equation}
\omega _{0}(z)dz:=\pi ^{-1}e^{-|z|^{2}}dz,  \tag{1.1}
\end{equation}%
$dz$ denotes the Lebesgue measure on $\mathbb{C=R}^{2},$ and may be defined
by the Rodrigues formula 
\begin{equation}
H_{m,n}(z,\overline{z})=(-1)^{m+n}e^{|z|^{2}}\partial _{z}^{n}\partial _{%
\overline{z}}^{m}(e^{-|z|^{2}}).  \tag{1.2}
\end{equation}

In random matrix theory, the measure $\omega _{0}(z)dz$ is closely related
to the so-called Ginibre model: a complex matrix whose entries are
independent standard Gaussian variables. Indeed, the eigenvalues of such a
matrix form a determinantal point process with respect to this measure (\cite%
{BF}). Letting the size of the Ginibre matrix tend to infinity, one obtains
the Ginibre determinantal point process whose kernel is $K_{0}(z,w)=e^{z%
\overline{w}}$. The latter is the reproducing kernel of Bargmann-Fock space $%
\mathcal{F}_{0}\left( \mathbb{C}\right) $ consisting of holomorphic
functions in the Hilbert space $\mathfrak{H}_{0}:=L^{2}(\mathbb{C},\omega
_{0}(z)dz)$. Beside, it turns out that $\mathcal{F}_{0}\left( \mathbb{C}%
\right) $ coincides with the null space 
\begin{equation}
\mathcal{A}_{0}\left( \mathbb{C}\right) :=\left\{ \phi \in \mathfrak{H}_{0},%
\text{ }\widetilde{\Delta }\phi =0\right\}  \tag{1.3}
\end{equation}%
of the second order differential operator 
\begin{equation}
\widetilde{\Delta }:=-\frac{\partial ^{2}}{\partial z\partial \overline{z}}+%
\overline{z}\frac{\partial }{\partial \overline{z}}  \tag{1.4}
\end{equation}%
called magnetic Laplacian \cite{AIM} and whose spectrum in $\mathfrak{H}_{0}$
consists of non negative integers $n=0,1,2,...$ . For any $n\geq 0$, an
orthogonal basis of the $n$-th eigenspace%
\begin{equation}
\mathcal{A}_{n}\left( \mathbb{C}\right) :=\left\{ \phi \in \mathfrak{H}_{0},%
\text{ }\widetilde{\Delta }\phi =n\phi \right\}  \tag{1.5}
\end{equation}%
is given by the complex Hermite polynomials $(H_{m,n}(z,\overline{z}%
))_{m\geq 0}$ and the corresponding reproducing kernel \cite{AIM} is
expressed as 
\begin{equation}
K_{n}(z,w)=K_{0}(z,w)L_{n}^{\left( 0\right) }(|z-w|),\text{ }z,w\in \mathbb{C%
}\text{,}  \tag{1.6}
\end{equation}%
where $L_{n}^{\left( 0\right) }$ is the $n$-th Laguerre polynomial. More
informatin on the $L^{2}$ spectral theory of $\widetilde{\Delta }$ can be
found in \cite{AM}. Since $K_{n}$ is the kernel of an orthogonal projection,
then it gives rise to a determinantal point process (\cite{Sos}, theorem 3,
p.12) referred to as Ginibre-type and studied in \cite{Shi},\cite{APRT} and 
\cite{MME}.

\subsection{\protect\medskip Deformation: the elliptic Ginibre model}

Let $\tau \in \lbrack 0,1)$ and let $U_{1},U_{2}$ be two independent
matrices drawn from the \textit{Gaussian Unitary Ensemble} (GUE). In this
respect, recall that the GUE consists of Hermitian matrices with Gaussian
independent entries. Then 
\begin{equation}
J_{\tau }:=\sqrt{1+\tau }U_{1}+i\sqrt{1-\tau }U_{2}  \tag{1.7}
\end{equation}%
is known as the elliptic Ginibre model \cite{BF} and interpolates between
the Ginibre model when $\tau =0$ and GUE in the degenerate limit as $\tau
\rightarrow 1^{-}$. Its eigenvalues process is again a determinantal point
process whose kernel is expressed through the rescaled Hermite polynomials 
\begin{equation}
H_{m}(z,\tau ):=\left( \frac{\tau }{2}\right) ^{m/2}H_{m}\left( \frac{z}{%
\sqrt{2\tau }}\right) ,m\geq 0,  \tag{1.8}
\end{equation}%
where the real Hermite polynomials 
\begin{equation}
H_{m}\left( x\right) =m!\sum\limits_{\ell =0}^{\left\lfloor \frac{1}{2}%
m\right\rfloor }\frac{\left( -1\right) ^{\ell }\left( 2x\right) ^{m-2\ell }}{%
\ell !\left( m-2\ell \right) !},\text{ }x\in \mathbb{R}\text{.}  \tag{1.9}
\end{equation}%
are chosen to be orthogonal on $\mathbb{R}$ with respect to the weight $%
e^{-x^{2}}$. The polynomials $H_{m}(z,\tau )$ are orthogonal with respect to
the elliptic measure 
\begin{equation}
\omega _{\tau }(z)d\nu \left( z\right) :=\pi ^{-1}e^{-(|z|^{2}-\tau \Re
(z^{2}))/(1-\tau ^{2})}dz.  \tag{1.10}
\end{equation}%
In particular, $H_{m}(z,0)=z^{m},$ $m\geq 0$ and $\left( H_{m}(z,\tau
)\right) _{m\geq 0}$ still belongs to the null (Bargmann-Fock) space of the
magnetic Laplacian $\widetilde{\Delta }$ for any $\tau \in \left[ 0,1\right) 
$. It is therefore natural to raise the following question: given a Landau
level $n\geq 1,$ what are the `natural' deformed analogues, say $\left(
H_{m,n}\left( z,\overline{z},\tau \right) \right) _{m\geq 0}$ of the complex
Hermite polynomials $\left( H_{m,n}\left( z,\overline{z}\right) \right)
_{m\geq 0}?.$By `natural', we here mean the sought polynomials must ensure
the commutativity of the following diagram:

\begin{equation}
\begin{array}{lcr}
H_{m,n}(z,\overline{z},\tau ) & \overset{\tau =0}{\longrightarrow } & 
H_{m,n}(z,\overline{z}) \\ 
. & . & . \\ 
n=0\big\downarrow & . & n=0\big\downarrow \\ 
. & . & . \\ 
H_{m}(z,\tau ) & \overset{\tau =0}{\longrightarrow } & z^{m}%
\end{array}%
,  \tag{1.11}
\end{equation}%
and must be $n$-polyanalytic for any fixed $n\geq 0$ as well. In other
words, they should satisfy 
\begin{equation}
\partial _{\overline{z}}^{n+1}H_{m,n}(z,\overline{z},\tau )=0,\quad m\geq 0,
\tag{1.12}
\end{equation}%
since belonging to the $n$-th eigenspace $\mathcal{A}_{n}\left( \mathbb{C}%
\right) $ of the operator $\widetilde{\Delta }.$

\medskip

By the virtue of $\left( 1.2\right) $, it is also tempting to define `%
\textit{\`{a} la Rodriguez}'\ the polynomials 
\begin{equation}
(-1)^{m+n}[\omega _{\tau }(z)]^{-1}\partial _{z}^{n}\partial _{\overline{z}%
}^{m}(\omega _{\tau }(z)),\quad m,n\geq 0,  \tag{1.13}
\end{equation}%
which clearly reduce to $H_{m,n}(z,\overline{z})$ when $\tau =0$. However, a
quick inspection shows that the polynomials in $\left( 1.13\right) $ are no
longer holomorphic at the lowest Landau level $n=0$ and for any $\tau \in
(0,1)$. Another attempt suggested by the expansion of the complex Hermite
polynomials 
\begin{equation}
H_{m,n}(z,\overline{z})=\sum_{k=0}^{m\wedge n}(-1)^{k}k!\binom{m}{k}\binom{n%
}{k}z^{m-k}\overline{z}^{n-k},  \tag{1.14}
\end{equation}%
\ \ \ \ \ \ \ \ \ \ \ \ \ \ \ \ \ \ \ \ \ \ \ \ \ \ \ \ \ \ \ \ \ \ \ \ \ \
\ \ \ \ \ \ \ \ \ \ \ \ \ \ \ \ \ \ \ \ \ \ \ \ \ \ \ \ \ \ \ \ \ \ \ \ \ \
\ \ \ \ \ \ \ \ \ \ \ \ \ \ \ \ \ \ \ \ \ \ \ \ \ \ \ \ \ \ \ \ \ \ \ \ \ \
\ \ \ \ \ \ \ \ \ \ \ \ \ \ \ \ \ \ \ \ \ \ \ \ \ \ \ \ \ \ \ \ \ \ \ \ \ \
\ \ \ \ \ \ \ \ \ \ \ \ \ \ \ \ \ \ \ \ \ \ \ \ \ \ \ \ \ \ \ \ \ \ \ \ \ \
\ \ \ \ \ \ \ \ \ \ \ \ \ \ \ \ \ \ \ \ \ \ \ \ \ \ \ \ \ \ \ \ \ \ \ \ \ \
\ \ \ \ \ \ \ \ \ \ \ \ \ \ \ \ \ \ \ \ \ \ \ \ \ \ \ \ \ \ \ \ \ \ \ \ \ \
\ \ \ \ \ \ \ \ \ \ \ \ \ \ \ \ \ \ \ \ \ \ \ \ \ \ \ \ \ \ \ \ \ \ \ \ \ \
\ \ \ \ \ \ \ \ \ \ \ \ \ \ \ \ \ \ \ \ \ \ \ \ \ \ \ \ \ \ \ \ \ \ \ \ \ \
\ \ \ \ \ \ \ \ \ \ \ \ \ \ \ \ \ \ \ \ \ \ \ \ \ \ \ \ \ \ \ \ \ \ \ \ \ \
\ \ \ \ \ \ \ \ \ \ \ \ \ \ \ \ \ \ \ \ \ \ \ \ \ \ \ \ \ \ \ \ \ \ \ \ \ \
\ \ \ \ \ \ \ \ \ \ \ \ \ \ \ \ \ \ \ \ \ \ \ \ \ \ \ \ \ \ \ \ \ \ \ \ \ \
\ \ \ \ \ \ \ \ \ \ \ \ \ \ \ \ \ \ \ \ \ \ \ \ \ \ \ \ \ \ \ \ \ \ \ \ \ \
\ \ \ \ \ \ \ \ \ \ \ \ \ \ \ \ \ \ \ \ \ \ \ \ \ \ \ \ \ \ \ \ \ \ \ \ \ \
\ \ \ \ \ \ \ \ \ \ \ \ \ \ \ \ \ \ \ \ \ \ \ \ \ \ \ \ \ \ \ \ \ \ \ \ \ \
\ \ \ \ \ \ \ \ \ \ \ \ \ \ \ \ \ \ \ \ \ \ \ \ \ \ \ \ \ \ \ \ \ \ \ \ \ \
\ \ \ \ \ \ \ \ \ \ \ \ \ \ \ \ \ \ \ \ \ \ \ \ \ \ \ \ \ \ \ \ \ \ \ \ \ \
\ \ \ \ \ \ \ \ \ \ \ \ \ \ \ \ \ \ \ \ \ \ \ \ \ \ \ \ \ \ \ \ \ \ \ \ \ \
\ \ \ \ \ \ \ \ \ \ \ \ \ \ \ \ \ \ \ \ \ \ \ \ \ \ \ \ \ \ \ \ \ \ \ \ \ \
\ \ \ \ \ \ \ \ \ \ \ \ \ \ \ \ \ \ \ \ \ \ \ \ \ \ \ \ \ \ \ \ \ \ \ \ \ \
\ \ \ \ \ \ \ \ \ \ \ \ \ \ \ \ \ \ \ \ \ \ \ \ \ \ \ \ \ \ \ \ \ \ \ \ \ \
\ \ \ \ \ \ \ \ \ \ \ \ \ \ \ \ \ \ \ \ \ \ \ \ \ \ \ \ \ \ \ \ \ \ \ \ \ \
\ \ \ \ \ \ \ \ \ \ \ \ \ \ \ \ \ \ \ \ \ \ \ \ \ \ \ \ \ \ \ \ \ \ \ \ \ \
\ \ \ \ \ \ \ \ \ \ \ \ \ \ \ \ \ \ \ \ \ \ \ \ \ \ \ \ \ \ \ \ \ \ \ \ \ \
\ \ \ \ \ \ \ \ \ \ \ \ \ \ \ \ \ \ \ \ \ \ \ \ \ \ \ \ \ \ \ \ \ \ \ \ \ \
\ \ \ \ \ \ \ \ \ \ \ \ \ \ \ \ \ \ \ \ \ \ \ \ \ \ \ \ \ \ \ \ \ \ \ \ \ \
\ \ \ \ \ \ \ \ \ \ \ \ \ \ \ \ \ \ \ \ \ \ \ \ \ \ \ \ \ \ \ \ \ \ \ \ \ \
\ \ \ \ \ \ \ \ \ \ \ \ \ \ \ \ \ \ \ \ \ \ \ \ \ \ \ \ \ \ \ \ \ \ \ \ \ \
\ \ \ \ \ \ \ \ \ \ \ \ \ \ \ \ \ \ \ \ \ \ \ \ \ \ \ \ \ \ \ \ \ \ \ \ \ \
\ \ \ \ \ \ \ \ \ \ \ \ \ \ \ \ \ \ \ \ \ \ \ \ \ \ \ \ \ \ \ \ \ \ \ \ \ \
\ \ \ \ \ \ \ \ \ \ \ \ \ \ \ \ \ \ \ \ \ \ \ \ \ \ \ \ \ \ \ \ \ \ \ \ \ \
\ \ \ \ \ \ \ \ \ \ \ \ \ \ \ \ \ \ \ \ \ \ \ \ \ \ \ \ \ \ \ \ \ \ \ \ \ \
\ \ \ \ \ \ \ \ \ \ \ \ \ \ \ \ \ \ \ \ \ \ \ \ \ \ \ \ \ \ \ \ \ \ \ \ \ \
\ \ \ \ \ \ \ \ \ \ \ \ \ \ \ would substitutes holomorphic monomials by
holomorphic Hermite polynomials as follows 
\begin{equation}
\sum_{k=0}^{m\wedge n}(-1)^{k}k!\binom{m}{k}\binom{n}{k}H_{m-k}(z,\tau )%
\overline{z}^{n-k},\quad m,n,\geq 0,  \tag{1.15}
\end{equation}%
which are obviously $n$-polyanalytic for any fixed $n\geq 0$. However, one
may prove for instance that the orthogonality with respect to $\omega _{\tau
}$ fails for $n=1$.

\bigskip \medskip

In order to find out the family of polynomials satisfying both the
orthogonality with respect to $\omega _{\tau },$ $\tau \in \left[ 0,1\right)
,$ and the poly-analyticity property, one needs to combine: $\left( i\right) 
$ the construction of the complex Hermite polynomials $H_{m,n}(z,\overline{z}%
)$ by means of creation and annihilation operators associated to the
magnetic Laplacian $\widetilde{\Delta }$ (see e.g., \cite{Shi}) with $\left(
ii\right) $ the construction of the holomorphic Hermite polynomials by means
of the deformed (squeezed) operators \cite{For-Jan}.

Actually, the first construction stems from the Heisenberg commutation
relation satisfied by creation and annihilation operators, which ensures
both needed properties (orthogonality and poly-analyticity). As to the
second construction, it exhibits squeezing of these operators while
preserving their commutation relation (this is known as the Bogoliubov
transformation). After careful computations, we obtain the following
representation of the deformed complex Hermite polynomials.

\medskip \textbf{Theorem 1.1. }\textit{For any Landau level }$n\geq 0,$%
\textit{\ an orthogonal basis of the eigenspace} 
\begin{equation}
\mathcal{A}_{n}\left( \mathbb{C}\right) :=\left\{ \phi \in L^{2}\left( 
\mathbb{C},\omega _{\tau }\left( z\right) dz\right) ,\widetilde{\Delta }\phi
=n\phi \right\}  \tag{1.16}
\end{equation}

\textit{is given by the following squeezed complex Hermite polynomials:}

\begin{equation}
H_{m,n}(z,\overline{z};\tau )=\frac{\left( \frac{\tau }{2}\right) ^{\left(
m+n\right) /2}}{\sqrt{m!}}\sum_{k=0}^{n\wedge m}i^{n-k}k!\binom{n}{k}\binom{m%
}{k}\frac{2^{k}\left( 1-\tau ^{2}\right) ^{k/2}}{\tau ^{k}}H_{m-k}\left( 
\frac{z}{\sqrt{2\tau }}\right) H_{n-k}\left( \frac{i\left( \overline{z}-\tau
z\right) }{\sqrt{2\tau \left( 1-\tau ^{2}\right) }}\right)  \tag{1.17}
\end{equation}

\smallskip

The formula $\left( 1.17\right) $ motivates the fact that $H_{m,n}(z,%
\overline{z};\tau )$ may be related to the so-called $2D$-Hermite
polynomials $H_{m,n}^{\left( R\right) }$ defined by their generating
function function \cite{DM}:%
\begin{equation}
\exp \left( -\frac{1}{2}\left\langle R\gamma ,\gamma \right\rangle
+\left\langle R\xi ,\gamma \right\rangle \right) =\sum\limits_{n,m=0}\frac{%
\gamma _{1}^{n}\gamma _{2}^{m}}{n!m!}H_{n,m}^{\left( R\right) }\left( \xi
_{1},\xi _{2}\right)  \tag{1.18}
\end{equation}%
where $R$ is a symmetric matrix, $\gamma =\left( \gamma _{1},\gamma
_{2}\right) \in \mathbb{C}^{2}$ and $\xi =\left( \xi _{1},\xi _{2}\right)
\in \mathbb{C}^{2}$. Indeed, it is the case and the following relation holds.

\medskip \textbf{Corollary 1.1}. \textit{For any }$m,n\geq 0,$\textit{\ the
polynomials }$H_{m,n}(z,\overline{z};\tau )$\textit{\ may also be expressed
in terms of the }$2D$\textit{-Hermite polynomial }$H_{n,m}^{\left( R_{\tau
}\right) }$ \textit{as}%
\begin{equation}
H_{m,n}(z,\overline{z};\tau )=\frac{i^{n}}{\sqrt{m!}}H_{n,m}^{\left( R_{\tau
}\right) }\left( \overline{z},\frac{i\left( \tau \overline{z}-z\right) }{%
\sqrt{1-\tau ^{2}}}\right)  \tag{1.19}
\end{equation}%
\textit{where }$R_{\tau }\in SL\left( 2,\mathbb{C}\right) $\textit{\ is the
symmetric matrix given by} 
\begin{equation}
R_{\tau }=\left( 
\begin{array}{cc}
\tau & i\sqrt{1-\tau ^{2}} \\ 
i\sqrt{1-\tau ^{2}} & \tau%
\end{array}%
\right) .  \tag{1.20}
\end{equation}

Another result we shall prove in this paper provides an isometric map $%
T_{\tau ,n}$ from the subspace $\mathcal{F}_{\tau }\left( \mathbb{C}\right) $
of entire functions in the Hilbert space $\mathfrak{H}_{\tau }:=L^{2}(%
\mathbb{C},\omega _{\tau }(z)dz)$ into the $n$-th eigenspace $\mathcal{A}%
_{n}\left( \mathbb{C}\right) $ of $\widetilde{\Delta }$ in $\mathfrak{H}%
_{0}. $ More precisely, this map is given by the following integral
transform.

\medskip \textbf{Proposition 1.1. }$T_{\tau ,n}:\mathcal{F}_{\tau }\left( 
\mathbb{C}\right) \rightarrow \mathcal{A}_{n}\left( \mathbb{C}\right) $ 
\textit{is defined by} 
\begin{equation}
T_{\tau ,n}\left[ \phi \right] \left( z\right) =\int\limits_{\mathbb{C}}%
\overline{W_{\tau ,n}\left( z,w\right) }\phi \left( w\right) \omega _{\tau
}(w)dw  \tag{1.21}
\end{equation}%
\textit{where the integral kernel is given by}%
\begin{equation}
W_{\tau ,n}\left( z,w\right) =\left( -\sqrt{\frac{\tau }{2}}\right) ^{n}%
\frac{e^{\overline{z}w-\tau \overline{z}^{2}/2}}{\sqrt{n!}}H_{n}\left( \frac{%
w-z}{\sqrt{2\tau }}-\sqrt{\frac{\tau }{2}}\overline{z}\right) .  \tag{1.22}
\end{equation}%
Using the symmetry relation $H_{n}\left( -x\right) =\left( -1\right)
^{n}H_{n}\left( x\right) ,$ $x\in \mathbb{R},$ the above kernel is
equivalently written \ as%
\begin{equation}
W_{\tau ,n}\left( z,w\right) =\left( \sqrt{\frac{\tau }{2}}\right) ^{n}\frac{%
e^{\overline{z}w-\tau \overline{z}^{2}/2}}{\sqrt{n!}}H_{n}\left( \sqrt{\frac{%
\tau }{2}}\overline{z}+\frac{z-w}{\sqrt{2\tau }}\right) .  \tag{1.23}
\end{equation}%
In particular,%
\begin{equation}
\lim_{\tau \rightarrow 0^{+}}W_{\tau ,n}\left( z,w\right) =\frac{e^{%
\overline{z}w}}{\sqrt{n!}}\left( z-w\right) ^{n}  \tag{1.24}
\end{equation}%
which is the kernel of the transform $T_{0,n}:$ $\mathcal{A}_{0}\left( 
\mathbb{C}\right) \rightarrow \mathcal{A}_{n}\left( \mathbb{C}\right) .$ \
For $\tau \in (0,1),$ this kernel $W_{\tau ,n}\left( z,w\right) $ is also
interpreted in terms of the two-photon coherent states \cite{YuE} and the
metaplectic representation \cite{ITZ} of the $SU(1,1)$ group.

The paper is organized as follows. In Section 2, we intoduce the squeezed
complex Hermite polynomials which will stand for the \textit{polyanalytic}
Hermite polynomials $H_{m,n}(z,\overline{z};\tau )$ associated with the
elliptic Ginibre model and we discuss their relation to the $2D$-Hermite
polynomials $H_{n,m}^{\left( R_{\tau }\right) }.$ Section 3 is devoted to
construct the integral transform $T_{\tau ,n}$ whose kernel will be
interpreted in terms of the two-photon coherent states.

\section{Squeezed complex Hermite polynomials}

\subsection{The magnetic picture: $\protect\tau =0$}

We shall revisit the construction of $H_{m,n}(z,\overline{z})$ by means of
the creation and annihilation operators associated with $\widetilde{\Delta }$%
. For this, we will follow the general scheme of the \textit{Fock
representation} for the Hamiltonian of the harmonic oscillator, see $\left( 
\text{\cite{Dirac}, pp.16-18}\right) $ for the general theory.

The Hamiltonian operator describing the dynamics of a particle of charge $e$
and mass $m_{\ast }$ on the Euclidean $xy$-plane, while interacting with a
perpendicular constant homogeneous magnetic field, is given by the operator%
\begin{equation}
H:=\frac{1}{2m_{\ast }}\left( i\hbar \nabla -\frac{e}{c}\mathbb{A}_{\nu
}\right) ^{2}  \tag{2.1}
\end{equation}%
where $\hbar $ denotes Planck's constant, $c$ is the light speed and $i$ the
imaginary unit. Denote by $\nu >0$ the strenght of the magnetic field $%
\mathbb{B}$ and select the symmetric gauge%
\begin{equation}
\mathbb{A}_{\nu }\mathbb{=-}\frac{1}{2}\mathbf{r\times }\mathbb{B}\mathbf{=}%
\left( -\frac{\nu }{2}y,\frac{\nu }{2}x\right)  \tag{2.2}
\end{equation}%
where $\mathbf{r}=\left( x,y\right) \in \mathbb{R}^{2}$. For simplicity, we
set $m_{\ast }=e=c=\hbar =1$ in $\left( 2.1\right) ,$ leading to the Landau
Hamiltonian%
\begin{equation}
H_{L}^{\nu }:=\frac{1}{2}\left( \left( i\partial _{x}-\frac{\nu }{2}y\right)
^{2}+\left( i\partial _{y}+\frac{\nu }{2}x\right) ^{2}\right)  \tag{2.3}
\end{equation}%
acting on the Hilbert space $L^{2}\left( \mathbb{R}^{2},dxdy\right) $. The
spectrum of $H_{L}^{\nu }$ consists of infinite number of eigenvalues with
infinite multiplicity of the form%
\begin{equation}
\epsilon _{n}^{\nu }:=\left( n+\frac{1}{2}\right) \nu ,\text{ }n=0,1,2,...%
\text{ .}  \tag{2.4}
\end{equation}%
We may intertwine the operator \ $H_{L}^{2\nu }$ by the unitary
transformation $Q:\phi \mapsto e^{\frac{1}{2}\nu |z|^{2}}\phi $ from $%
L^{2}\left( \mathbb{R}^{2},dxdy\right) $ into $L^{2}\left( \mathbb{C}%
,e^{-\nu |z|^{2}}dz\right) $, the so-called ground state transformation as
follows%
\begin{equation}
\Delta _{\nu }:=e^{\frac{1}{2}\nu |z|^{2}}\left( \frac{1}{2}H_{L}^{2\nu }-%
\frac{\nu }{2}\right) e^{-\frac{1}{2}\nu |z|^{2}}=-\frac{\partial ^{2}}{%
\partial z\partial \overline{z}}+\nu \overline{z}\frac{\partial }{\partial 
\overline{z}}  \tag{2.5}
\end{equation}%
For $\nu =1,\Delta _{1}=\widetilde{\Delta }$ in $\left( 2.5\right) $ and

\begin{equation}
H_{L}:=\frac{1}{2}H_{L}^{2\nu }-\frac{\nu }{2}=\frac{1}{2}\left( \left[ -%
\frac{1}{2}\left( \left( \partial _{x}+\frac{i}{2}y\right) ^{2}+\left(
\partial _{y}-\frac{i}{2}x\right) ^{2}\right) \right] -1\right)  \tag{2.6}
\end{equation}%
which decomposes as 
\begin{equation}
H_{L}=:A^{\ast }A:=\left( \partial _{z}-\frac{1}{2}\overline{z}\right)
\left( -\partial _{\overline{z}}-\frac{1}{2}z\right) .  \tag{2.7}
\end{equation}%
The annihilation $A$ and the creation operators $A^{\star }$ satisfy the
Heisenberg commutation relation $[A,A^{\star }]=1$. Yet, the following
operators also do: 
\begin{equation}
B=-\partial _{z}-\frac{1}{2}\overline{z},\quad B^{\star }=\partial _{%
\overline{z}}-\frac{1}{2}z,  \tag{2.8}
\end{equation}%
\ That is $[B,B^{\star }]=1$, and commute with\ $A$ and $A^{\star }$ as
well. \ As explained in \cite{Ali}, the essence of the operators $\left(
B,B^{\star }\right) $ stems from the change of orientation of the magnetic
vector field along the direction orthogonal to the plane $\mathbb{R}%
^{2}\subset \mathbb{R}^{3}$, or equivalently to the choice of the vector
potential $(y,-x,0)$. At the mathematical level, the (commuting) algebras $%
\{A,A^{\star }\}$ and $\{B,B^{\star }\}$ reflect the complex conjugation $%
z\mapsto \overline{z}$ (holomorphic and anti-holomorphic structures) and the
two choices of the vector potentials reflect the complex structures $%
z\mapsto \pm iz$.

\smallskip Now, let%
\begin{equation*}
\psi _{0}(z,\overline{z}):=\pi ^{-\frac{1}{2}}e^{-\frac{1}{2}|z|^{2}},
\end{equation*}%
then $A\psi _{0}=B\psi _{0}=0$ and the eigenstates of $H_{L}$ corresponding
to the lowest Landau level $n=0,$ are given by (\cite{For-Jan}): 
\begin{equation}
\psi _{m}^{(0)}(z,\overline{z})=\frac{1}{\sqrt{m!}}(-B^{\star })^{m}\psi
_{0}(z,\overline{z})=\frac{z^{m}}{\sqrt{\pi m!}}e^{-\frac{1}{2}%
|z|^{2}},\quad m\geq 0.  \tag{2.9}
\end{equation}%
More generally, the eigenstates corresponding to highest Landau levels $%
n\geq 0$ are given by (\cite{Shi}): 
\begin{equation}
\psi _{m}^{(n)}(z,\overline{z})=\frac{1}{\sqrt{n!}}(A^{\star })^{n}\psi
_{m}^{(0)}(z,\overline{z})=H_{m,n}(z,\overline{z})e^{-\frac{1}{2}|z|^{2}},%
\text{ \ \ }m,n\geq 0.  \tag{2.10}
\end{equation}%
This formula stems from the commutation relation $[H,A^{\star }]=A^{\star }$
which allows to come from the $n$-th Landau level to the $(n+1)$-th one.
Keeping in mind the invertible transformation $Q$ above, one obtains the
spectral resolution of $\widetilde{\Delta }$ in $L^{2}\left( \mathbb{C}%
,\omega _{0}(z)dz\right) \mathbf{.}$

\subsection{The deformed setting $\protect\tau \in \left[ 0,1\right) $:
squeezing the lowest Landau level}

In this paragraph, we briefly recall the construction of the holomorphic
Hermite polynomials using squeezed creation and annihilation operators. Our
presentation is taken from \cite{For-Jan}. For a fixed radiation mode with
photon annihilation operator $B,$ let us consider the Bogoliubov
transformation

\begin{equation}
\left\{ 
\begin{array}{c}
B_{\mu }=(\cosh \mu )B-(\sinh \mu )B^{\star } \\ 
B_{\mu }^{\star }=-(\sinh \mu )B+(\cosh \mu )B^{\star }%
\end{array}%
\right.  \tag{2.11}
\end{equation}%
From $\cosh ^{2}\mu -\sinh ^{2}\mu =1,$ it follows that%
\begin{equation}
\left[ B_{\mu },B_{\mu }^{\ast }\right] =1  \tag{2.12}
\end{equation}%
which means that the transformation $\left( 2.11\right) $ leaves the
commuator $\left[ B,B^{\ast }\right] $ invariant. Thus, the change of
variables from $\left( B,B^{\ast }\right) $ to $\left( B_{\mu },B_{\mu
}^{\ast }\right) $ is a linear cononical transformation \cite{Wolf}. To find
out the ground state one has to solve the equation 
\begin{equation}
B_{\mu }\psi =0.  \tag{2.13}
\end{equation}%
A solution to this equation is given by 
\begin{equation}
\psi _{\mu }(z,\overline{z})=\frac{1}{\sqrt{\pi }}e^{-\frac{1}{2}%
(|z|^{2}-\tau z^{2})},\text{ \ \ \ }\tau =\tanh \mu .  \tag{2.14}
\end{equation}%
From the latter one, we propose the weight function 
\begin{equation}
\omega _{\tau }(z)dz=\left\vert \psi _{\mu }\left( \frac{z}{\sqrt{1-\tau ^{2}%
}},\frac{\overline{z}}{\sqrt{1-\tau ^{2}}}\right) \right\vert ^{2}\frac{dz}{%
\sqrt{1-\tau ^{2}}}.  \tag{2.15}
\end{equation}%
Moreover, the deformed analogue of $\left( 2.1\right) $ is $\left( \cite%
{For-Jan}\right) :$ 
\begin{equation}
\frac{1}{\sqrt{m!}}(-B_{\mu }^{\star })^{m}\psi _{\mu }(z,\overline{z})=%
\frac{1}{\sqrt{m!}}\left( \frac{\tanh \mu }{2}\right) ^{m/2}H_{m}\left( 
\frac{z}{\sqrt{\sinh 2\mu }}\right) \psi _{\mu }(z,\overline{z}),\quad m\geq
0.  \tag{2.16}
\end{equation}%
Note in passing that with the substitutions 
\begin{equation}
\tau =\tanh \mu ,\text{ \ }z\mapsto z\cosh \mu =\frac{z}{\sqrt{1-\tau ^{2}}},
\tag{2.17}
\end{equation}%
the right hand side of $\left( 2.16\right) $ transforms into: 
\begin{equation*}
\frac{1}{\sqrt{m!}}\left( \frac{\tau }{2}\right) ^{m/2}H_{m}\left( \frac{z}{%
\sqrt{2\tau }}\right) \psi _{\mu }\left( \frac{z}{\sqrt{1-\tau ^{2}}},\frac{%
\overline{z}}{\sqrt{1-\tau ^{2}}}\right)
\end{equation*}

\begin{equation}
=\frac{1}{\sqrt{m!}}H_{m}(z,\tau )\psi _{\mu }\left( \frac{z}{\sqrt{1-\tau
^{2}}},\frac{\overline{z}}{\sqrt{1-\tau ^{2}}}\right)  \tag{2.18}
\end{equation}%
Before proceeding to the definition of the squeezed complex Hermite
polynomials, we would like to stress that though $H_{m}\left( z,\tau \right) 
$ were identified in \cite{For-Jan} through their recurrence relation, one
may rather use the Rodrigues formula for the Hermite polynomials. Indeed,
the squeezed creation operator $-B_{\mu }^{\ast }$ may be written as 
\begin{equation}
-B_{\mu }^{\ast }=(\sinh \mu )B-(\cosh \mu )B^{\ast }  \tag{2.19}
\end{equation}

\begin{equation}
=-\left( (\sinh \mu )\left( \partial _{z}+\frac{\overline{z}}{2}\right)
+\left( \cosh \mu \right) \left( \partial _{\overline{z}}+\frac{z}{2}\right)
-z\cosh \mu \right)  \tag{2.20}
\end{equation}%
\begin{equation}
=-e^{-z\overline{z}/2}\left( \left( \sinh \mu \right) \partial _{z}+\left(
\cosh \mu \right) \partial _{\overline{z}}-z\cosh \mu \right) e^{z\overline{z%
}/2}.  \tag{2.21}
\end{equation}%
Whence we deduce that

\begin{equation}
\left( -B_{\mu }^{\ast }\right) ^{m}\left[ \psi _{\mu }\right] \left( z,%
\overline{z}\right) =\left( -1\right) ^{m}e^{-z\overline{z}/2}\left( (\sinh
\mu )\partial _{z}+(\cosh \mu )\partial _{\overline{z}}-z\cosh \mu \right)
^{m}\left[ e^{\frac{1}{2}z^{2}\tanh \mu }\right]  \tag{2.22}
\end{equation}%
\begin{equation}
=\left( -1\right) ^{m}e^{-z\overline{z}/2}\left( (\sinh \mu )\partial
_{z}-z\cosh \mu \right) ^{m}\left[ e^{\frac{1}{2}z^{2}\tanh \mu }\right] 
\tag{2.23}
\end{equation}%
\begin{equation}
=\left( -\sinh \mu \right) ^{m}e^{-z\overline{z}/2}\left( \partial
_{z}-z\cosh \mu \right) ^{m}\left[ e^{\frac{1}{2}z^{2}\tanh \mu }\right] , 
\tag{2.24}
\end{equation}%
where $\left( 2.20\right) $ being a consequence of the holomorphy of $%
z\mapsto e^{\frac{1}{2}z^{2}\tanh \mu }.$ Noting further that 
\begin{equation}
\partial _{z}-z\cosh \mu =e^{\frac{1}{2}z^{2}\coth \mu }\partial _{z}\left(
e^{-\frac{1}{2}z^{2}\coth \mu }\right) ,  \tag{2.25}
\end{equation}%
and using the Rodrigues formula for the Hermite polynomials, it follows that 
\begin{equation}
\left( -B_{\mu }^{\ast }\right) ^{m}\left[ \psi _{\mu }\right] \left( z,%
\overline{z}\right) =\left( -\sinh \mu \right) ^{m}e^{-z\overline{z}/2}e^{%
\frac{1}{2}z^{2}\coth \mu }\partial _{z}\left( e^{-z^{2}/\sinh 2\mu }\right)
\tag{2.26}
\end{equation}%
\begin{equation}
=\frac{\left( \sinh \mu \right) ^{m}}{\left( \sinh 2\mu \right) ^{m/2}}e^{-z%
\overline{z}/2}e^{\frac{1}{2}z^{2}\coth \mu }e^{-z^{2}/\sinh 2\mu
}H_{m}\left( \frac{z}{\sqrt{\sinh 2\mu }}\right)  \tag{2.27}
\end{equation}

\begin{equation}
=\left( \frac{\tanh \mu }{2}\right) ^{m/2}H_{m}\left( \frac{z}{\sqrt{\sinh
2\mu }}\right) \psi _{\mu }\left( z,\overline{z}\right)  \tag{2.28}
\end{equation}%
as in $\left( 2.16\right) .$

\subsection{Defining and computing the squeezed complex Hermite polynomials}

By mimiking $\left( 2.10\right) ,$ the commutation relation $\left[
H_{L},A^{\ast }\right] =A^{\ast }$ motivates the following.

\textbf{Definition 2.1. }\textit{For any} $n\geq 0,$ \textit{let} 
\begin{equation}
\psi _{m}^{(n)}(z,\overline{z},\tanh \mu ):=\frac{1}{\sqrt{n!}}(A^{\star
})^{n}\left[ \frac{1}{\sqrt{m!}}\left( \frac{\tanh \mu }{2}\right)
^{m/2}H_{m}\left( \frac{z}{\sqrt{\sinh 2\mu }}\right) \psi _{\mu }(z,%
\overline{z})\right] .  \tag{2.29}
\end{equation}%
In this respect, the following proposition is a major step toward the proof
of Theorem 1.1.

\textbf{Proposition 2.1. }\textit{For any}\textbf{\ }$n\geq 0,\left( \psi
_{m}^{(n)}(z,\overline{z},\tanh \mu )\right) _{m\geq 0}$ \textit{are
orthogonal eigenfunctions of }$H_{L}$\textit{\ in }$L^{2}\left( \mathbb{C}%
,dz\right) ,$\textit{\ corresponding to the }$n$\textit{th Landau level}%
.\medskip

\textit{Proof.} By the virtue of the commutation relation $[H,A^{\star
}]=A^{\star }$, the family $(\psi _{m}^{(n)}(z,\overline{z},\tanh \mu
))_{m\geq 0}$ belongs to the $n$-th eigenspace of $H_{L}$ in $L^{2}(\mathbb{C%
},dz)$. As to the orthogonality property, it follows from that of $(\psi
_{m}^{(0)}(z,\overline{z},\tanh \mu ))_{m\geq 0}$ or equivalently of the
holomorphic Hermite polynomials with respect to $\omega _{\tau }(z)dz$ .
Indeed, $A$ and $A^{\star }$ are adjoint to each other in $L^{2}(\mathbb{C}%
,dz)$ and Lemma 2.2 in \cite{Shi} shows that 
\begin{equation}
A^{n}(A^{\star })^{n}=\prod_{k=1}^{n}[H+k].  \tag{2.30}
\end{equation}%
Consequently, for any $m,m^{\prime }\geq 0$, 
\begin{equation}
\langle \psi _{m}^{(n)},\psi _{m^{\prime }}^{(n)}\rangle _{L^{2}(\mathbb{C}%
,dz)}=\langle (A^{n}(a^{\star })^{n}\psi _{m}^{(0)},\psi _{m^{\prime
}}^{(0)}\rangle _{L^{2}(\mathbb{C},dz)}=n!\delta _{mm^{\prime }},  \tag{2.31}
\end{equation}%
as claimed.

\medskip

We now are ready to prove Theorem 1.1. Indeed, $(A^{\star })^{n}$ is a
differential operator with polynomial coefficients in $(z,\overline{z})$ and
as such, 
\begin{equation}
\psi _{m}^{(n)}(z,\overline{z},\tanh \mu )=G_{m,n}(z,\overline{z},\tanh \mu
)\psi _{\mu }(z,\overline{z}),  \tag{2.32}
\end{equation}%
for some polynomials $G_{m,n}(z,\overline{z},\tanh \mu )$. Up to a
rescaling, these polynomials are polyanalytic analogues of the holomorphic
Hermite polynomials $H_{m}(z,\tau =\tanh \mu )$ for the higher Landau levels 
$n\geq 1$, or equivalently the squeezed complex Hermite polynomials.

\bigskip

\subsection{Polyanalytic Hermite polynomials associated with the elliptic
Ginibre model}

In this paragraph, we shall prove the following result:

\medskip

\textbf{Proposition 2.2.} \textit{For any }$m,n\geq 0$, 
\begin{equation}
G_{m,n}(z,\overline{z},\tanh (\mu ))=\frac{1}{\sqrt{m!}}\left( \frac{\tanh
(\mu )}{2}\right) ^{m/2}\sum_{k=0}^{n\wedge m}k!\binom{n}{k}\binom{m}{k}%
\frac{2^{k}}{\sinh ^{k/2}(2\mu )}\left( i\sqrt{\frac{\tanh (\mu )}{2}}%
\right) ^{n-k}  \tag{2.33}
\end{equation}%
\begin{equation*}
\times H_{m-k}\left( \frac{z}{\sqrt{\sinh (2\mu )}}\right) H_{n-k}\left( -i%
\sqrt{\frac{\tanh (\mu )}{2}}z+i\frac{\overline{z}}{\sqrt{2\tanh (\mu )}}%
\right)
\end{equation*}%
\textbf{Proof.} Recall that $A^{\star }=\partial _{z}-(\overline{z}/2)$.
Since $[\partial _{z},\overline{z}]=0,$ then 
\begin{equation}
(A^{\star })^{n}=\sum_{j=0}^{n}\binom{n}{j}\left( -\frac{\overline{z}}{2}%
\right) ^{n-j}\partial _{z}^{j}.  \tag{2.34}
\end{equation}%
Using the differential relation $\partial _{z}H_{m}(z)=2mH_{m-1}(z)$
together with Leibniz formula, we get for any $0\leq j\leq n$, 
\begin{equation}
\partial _{z}^{j}\left[ H_{m}\left( \frac{z}{\sqrt{\sinh 2\mu }}\right) \psi
_{\mu }(z,\overline{z})\right] =\sum_{k=0}^{j}\binom{j}{k}\frac{2^{k}m!}{%
\sinh ^{k/2}(2\mu )(m-k)!}H_{m-k}\left( \frac{z}{\sqrt{\sinh 2\mu }}\right)
\partial _{z}^{j-k}\psi _{\mu }(z,\overline{z}),  \tag{2.35}
\end{equation}%
where $H_{m-k}=0$ if $k>m$. Consequently, we, successively, get

\begin{equation}
(A^{\star })^{n}\left[ H_{m}\left( \frac{z}{\sqrt{\sinh 2\mu }}\right) \psi
_{\mu }(z,\overline{z})\right]  \tag{2.36}
\end{equation}

\begin{equation}
=\sum_{j=0}^{n}\binom{n}{j}\left( -\frac{\overline{z}}{2}\right)
^{n-j}\sum_{k=0}^{j}\binom{j}{k}\frac{2^{k}m!}{(m-k)!\sinh ^{k/2}2\mu }%
H_{m-k}\left( \frac{z}{\sqrt{\sinh 2\mu }}\right) \partial _{z}^{j-k}\psi
_{\mu }(z,\overline{z})  \tag{2.37}
\end{equation}%
\begin{equation}
=\sum_{k=0}^{n\wedge m}\binom{m}{k}\frac{2^{k}}{\sinh ^{k/2}2\mu }%
H_{m-k}\left( \frac{z}{\sqrt{\sinh 2\mu }}\right) \sum_{j=k}^{n}\frac{n!}{%
(n-j)!(j-k)!}\left( -\frac{\overline{z}}{2}\right) ^{n-j}\partial
_{z}^{j-k}\psi _{\mu }(z,\overline{z})  \tag{2.38}
\end{equation}%
\begin{equation}
=\sum_{k=0}^{n\wedge m}k!\binom{n}{k}\binom{m}{k}\frac{2^{k}}{\sinh
^{k/2}2\mu }H_{m-k}\left( \frac{z}{\sqrt{\sinh 2\mu }}\right)
\sum_{j=0}^{n-k}\binom{n-k}{j}\left( -\frac{\overline{z}}{2}\right)
^{n-k-j}\partial _{z}^{j}\psi _{\mu }(z,\overline{z}).  \tag{2.39}
\end{equation}%
Next, we infer from \cite{Ben-Gha} that%
\begin{equation}
\partial _{z}^{j}\psi _{\mu }(z,\overline{z})=(-1)^{j}I_{j}^{\frac{1}{2},%
\frac{1}{2}\tanh \mu }(z,\overline{z})\psi _{\mu }(z,\overline{z}) 
\tag{2.40}
\end{equation}%
where 
\begin{equation}
I_{j}^{\frac{1}{2},\frac{1}{2}\tanh \mu }(z,\overline{z})=I_{j}^{\frac{1}{2},%
\frac{1}{2}\tanh \mu }(z,\overline{z}|0)  \tag{2.41}
\end{equation}%
in the notations of that paper $\left( \text{see eq.}\left( 1.3\right)
\right) $. Actually, Corollay 2.9 there shows that 
\begin{equation}
(-1)^{j}I_{j}^{\frac{1}{2},\frac{1}{2}\tanh \mu }(z,\overline{z})=\left( i%
\sqrt{\tanh \mu }/\sqrt{2}\right) ^{j}H_{j}\left( \frac{z\tanh \mu -%
\overline{z}/2}{i\sqrt{2\tanh \mu }}\right) .  \tag{2.42}
\end{equation}%
As a result, we obtain%
\begin{equation*}
(A^{\star })^{n}\left[ H_{m}\left( \frac{z}{\sqrt{\sinh 2\mu }}\right) \psi
_{\mu }(z,\overline{z})\right] =\psi _{\mu }(z,\overline{z}%
)\sum_{k=0}^{n\wedge m}k!\binom{n}{k}\binom{m}{k}\frac{2^{k}}{\sinh
^{k/2}2\mu }H_{m-k}\left( \frac{z}{\sqrt{\sinh 2\mu }}\right)
\end{equation*}%
\begin{equation}
\times \sum_{j=0}^{n-k}\binom{n-k}{j}\left( -\frac{\overline{z}}{2}\right)
^{n-k-j}[i\sqrt{\tanh \mu }/\sqrt{2}]^{j}H_{j}\left( \frac{z(\tanh \mu )-%
\overline{z}/2}{i\sqrt{2\tanh \mu }}\right) .  \tag{2.43}
\end{equation}%
Applying the Taylor expansion, for the Hermite polynomials%
\begin{equation}
H_{n-k}\left( u+t\right) =\sum\limits_{j=0}^{n-k}\left( 
\begin{array}{c}
n-k \\ 
j%
\end{array}%
\right) \left( 2u\right) ^{n-k-j}H_{j}\left( t\right) ,  \tag{2.44}
\end{equation}%
to the inner sum, we further get 
\begin{equation*}
(A^{\star })^{n}\left[ H_{m}\left( \frac{z}{\sqrt{\sinh 2\mu }}\right) \psi
_{\mu }(z,\overline{z})\right] =\psi _{\mu }(z,\overline{z}%
)\sum_{k=0}^{n\wedge m}k!\binom{n}{k}\binom{m}{k}\frac{2^{k}}{\sinh
^{k/2}2\mu }
\end{equation*}%
\begin{equation}
\left( i\sqrt{\frac{\tanh \mu }{2}}\right) ^{n-k}H_{m-k}\left( \frac{z}{%
\sqrt{\sinh 2\mu }}\right) H_{n-k}\left( -i\sqrt{\frac{\tanh \mu }{2}}z+i%
\frac{\overline{z}}{\sqrt{2\tanh \mu }}\right)  \tag{2.45}
\end{equation}%
Keeping in mind $\left( 2.32\right) $, the announced expression of $%
G_{mn.n}(z,\overline{z},\tanh \mu )$ follows.

\medskip In this expression, we again substitute 
\begin{equation}
\tanh \mu \rightarrow \tau ,\text{ }z\mapsto z/\sqrt{1-\tau ^{2}}=z\cosh \mu
,  \tag{2.46}
\end{equation}%
and get the expression of $H_{m,n}(z,\overline{z},\tau )$ displayed in
Theorem 1.1, with the form

\begin{equation}
H_{m,n}(z,\overline{z},\tau ):=\frac{1}{\sqrt{m!}}\left( \frac{\tau }{2}%
\right) ^{(m+n)/2}\sum_{k=0}^{n\wedge m}i^{n-k}k!\binom{n}{k}\binom{m}{k}%
\frac{2^{k}(1-\tau ^{2})^{k/2}}{\tau ^{k}}  \tag{2.47}
\end{equation}%
\begin{equation*}
\times H_{m-k}\left( \frac{z}{\sqrt{2\tau }}\right) H_{n-k}\left( \frac{i(%
\overline{z}-\tau z)}{\sqrt{2\tau (1-\tau ^{2})}}\right) .
\end{equation*}%
\textbf{Remark 2.1}. Letting $\tau \rightarrow 0^{+}$, we have%
\begin{equation}
\lim_{\tau \rightarrow 0^{+}}\left( \frac{\tau }{2}\right) ^{m/2}\frac{%
2^{k/2}(1-\tau ^{2})^{k/2}}{\tau ^{k/2}}H_{m-k}\left( \frac{z}{\sqrt{2\tau }}%
\right) =z^{m-k},  \tag{2.48}
\end{equation}%
and similarly 
\begin{equation}
\lim_{\tau \rightarrow 0^{+}}\left( \frac{\tau }{2}\right)
^{(n-k)/2}H_{n-k}\left( -i\sqrt{\frac{\tau }{2(1-\tau ^{2})}}z+\frac{i}{%
\sqrt{2\tau (1-\tau ^{2})}}\overline{z}\right) =i^{n-k}\overline{z}^{n-k}. 
\tag{2.49}
\end{equation}%
It follows that 
\begin{equation}
\lim_{\tau \rightarrow 0^{+}}H_{m,n}(z,\overline{z},\tau )=(-1)^{n}H_{m,n}(z,%
\overline{z}).  \tag{2.50}
\end{equation}%
If we discard the factor $1/\sqrt{m!}$ and use the generating function for
Hermite polynomials, then we readily derive 
\begin{equation}
\sum_{m,n\geq 0}\sqrt{m!}H_{m,n}(z,\overline{z},\tau )\frac{u^{m}}{m!}\frac{%
v^{n}}{n!}=\exp \left( \tau (v^{2}-u^{2})/2+uz-v(\overline{z}-\tau z)/\sqrt{%
1-\tau ^{2}}+uv\sqrt{1-\tau ^{2}}\right) .  \tag{2.51}
\end{equation}%
In particular, we recover the known result (see for instance \cite{Ism-Zha},
eq. (3.1)): 
\begin{equation}
\sum_{m,n\geq 0}\sqrt{m!}H_{m,n}(z,\overline{z},0)\frac{u^{m}}{m!}\frac{%
(-v)^{n}}{n!}=e^{uz+v\overline{z}-uv}.  \tag{2.52}
\end{equation}%
\textbf{Remark 2.2. }From the\textbf{\ }property $H_{n-k}\left( u\right)
=\left( -1\right) ^{n-k}H_{n-k}\left( -u\right) $\textbf{, }one readly gets
the relation\textbf{\ }$\overline{H_{m,n}(z,\overline{z},\tau )}=H_{m,n}(%
\overline{z},z,\tau ).$

\subsection{Relation to $2D$-Hermite polynomials$:$ proof of Corollary 1.2}

Following \cite{DM} the two-dimensional Hermite polynomials $H_{n,m}^{\left(
R\right) }\left( \xi _{1},\xi _{2}\right) $ \ associated to a symmetric
matrix \ $R=\left( r_{kl}\right) _{1\leq l,k\leq 2}$ are defined by the
following generating function 
\begin{equation}
\exp \left( -\frac{1}{2}\left\langle R\gamma ,\gamma \right\rangle
+\left\langle R\xi ,\gamma \right\rangle \right) =\sum\limits_{n,m=0}\frac{%
\gamma _{1}^{n}\gamma _{2}^{m}}{n!m!}H_{n,m}^{\left( R\right) }\left( \xi
_{1},\xi _{2}\right)  \tag{2.53}
\end{equation}%
where $\xi =\left( \xi _{1},\xi _{2}\right) \in $ $\mathbb{C}^{2}$ , $\gamma
=\left( \gamma _{1},\gamma _{2}\right) \in \mathbb{C}^{2},\xi _{1},\xi _{2},$
$\gamma _{1},\gamma _{2}$ are arbitrary complex numbers and$\ \left\langle
R\gamma ,\gamma \right\rangle =\sum\limits_{j,k=1}^{2}\gamma
_{j}r_{jk}\gamma _{k}.$ These polynomials may be expressed through one
variable Hermite polynomials as $\left( \cite{DM}\text{, eq.8}\right) :$

\begin{equation}
H_{n,m}^{\left( R\right) }\left( \xi _{1},\xi _{2}\right) =\left( \frac{%
r_{11}^{n}r_{22}^{m}}{2^{n+m}}\right) ^{\frac{1}{2}}\sum\limits_{k=0}^{n%
\wedge m}\left( \frac{-2r_{12}}{\sqrt{r_{11}r_{22}}}\right) ^{k}\frac{n!m!}{%
\left( n-k\right) !\left( m-k\right) !k!}H_{n-k}\left( \frac{\zeta _{1}}{%
\sqrt{2r_{11}}}\right) H_{m-k}\left( \frac{\zeta _{2}}{\sqrt{2r_{22}}}\right)
\tag{2.54}
\end{equation}%
where $\zeta =\left( \zeta _{1},\zeta _{2}\right) \in \mathbb{C}^{2}$ is
connected to $\xi =\left( \xi _{1},\xi _{2}\right) $ by 
\begin{equation}
\left( 
\begin{array}{cc}
r_{11} & r_{12} \\ 
r_{12} & r_{22}%
\end{array}%
\right) \left( 
\begin{array}{c}
\xi _{1} \\ 
\xi _{2}%
\end{array}%
\right) =\left( 
\begin{array}{c}
\zeta _{1} \\ 
\zeta _{2}%
\end{array}%
\right) .  \tag{2.55}
\end{equation}%
By the virtue of Theorem 1.1, we are led to consider the unimodular
symmetric matrix $R=R_{\tau }\in \mathcal{S}L\left( 2,\mathbb{C}\right) ,$
whose entries are given by 
\begin{equation}
r_{11}=r_{22}=\tau ,\text{ \ \ }r_{12}=r_{21}=i\sqrt{1-\tau ^{2}}, 
\tag{2.56}
\end{equation}%
in which case the $2D$-Hermite polynomials read 
\begin{equation}
H_{n,m}^{\left( R\right) }\left( \xi _{1},\xi _{2}\right) =\left( \frac{\tau 
}{2}\right) ^{\frac{1}{2}\left( n+m\right) }\sum\limits_{k=0}^{n\wedge
m}\left( \frac{2\sqrt{1-\tau ^{2}}}{i\tau }\right) ^{k}k!\left( 
\begin{array}{c}
n \\ 
k%
\end{array}%
\right) \left( 
\begin{array}{c}
m \\ 
k%
\end{array}%
\right) H_{m-k}\left( \frac{\zeta _{1}}{\sqrt{2\tau }}\right) H_{n-k}\left( 
\frac{\zeta _{2}}{\sqrt{2\tau }}\right)  \tag{2.57}
\end{equation}%
Consequently, we may choose $\zeta _{1}$ and $\zeta _{2}$ as 
\begin{equation}
\zeta _{1}=z,\text{ }\zeta _{2}=\frac{i\left( \overline{z}-\tau z\right) }{%
\sqrt{1-\tau ^{2}}}  \tag{2.58}
\end{equation}%
or equivalently, by virtue of $\left( 2.55\right) :$%
\begin{equation}
\left( 
\begin{array}{c}
\xi _{1} \\ 
\xi _{2}%
\end{array}%
\right) =R_{\tau }^{-1}\left( 
\begin{array}{c}
\zeta _{1} \\ 
\zeta _{2}%
\end{array}%
\right) =\left( 
\begin{array}{cc}
\tau & -i\sqrt{1-\tau ^{2}} \\ 
-i\sqrt{1-\tau ^{2}} & \tau%
\end{array}%
\right) \left( 
\begin{array}{c}
z \\ 
\frac{i\left( \overline{z}-\tau z\right) }{\sqrt{1-\tau ^{2}}}%
\end{array}%
\right) =\left( 
\begin{array}{c}
\overline{z} \\ 
\frac{i\left( \tau \overline{z}-z\right) }{\sqrt{1-\tau ^{2}}}%
\end{array}%
\right)  \tag{2.59}
\end{equation}%
Taking into account the factor $i^{n}/\sqrt{m!}$the corollary is proved.

\medskip

\section{An integral transform $T_{\protect\tau ,m}:\mathcal{F}_{\protect%
\tau }\left( \mathbb{C}\right) \rightarrow \mathcal{A}_{n}\left( \mathbb{C}%
\right) $}

\subsection{The kernel $W_{\protect\tau ,n}\left( z,w\right) $ of $T_{%
\protect\tau ,m}$}

Let $\mathcal{F}_{\tau }\left( \mathbb{C}\right) $ be the space of entire
functions in the Hilbert space $\mathfrak{H}_{\tau }:=L^{2}(\mathbb{C}%
,\omega _{\tau }(z)dz).$ An orthogonal basis of this space is given by the
normalized holomorphic Hermite polynomials $\left( \cite{Byu-For},\text{ p.21%
}\right) $ :

\begin{equation}
h_{\tau ,k}\left( w\right) =\frac{1}{\sqrt{k!}}\left( \frac{\tau }{2}\right)
^{k/2}H_{k}\left( \frac{w}{\sqrt{2\tau }}\right) ,\text{ }k\geq 0.  \tag{3.1}
\end{equation}%
Let us recall $\left( 1.16\right) $ where we denoted by $\mathcal{A}%
_{n}\left( \mathbb{C}\right) $ the $n$-th polyanalytic eigenspace of the
operator $\widetilde{\Delta }$ in $\mathfrak{H}_{\tau },$ a basis of which
consists of the normalized complex Hermite polynomials%
\begin{equation}
\phi _{m,n}\left( z,\overline{z}\right) :=\frac{H_{m,n}\left( z,\overline{z}%
\right) }{\sqrt{m!n!}}  \tag{3.2}
\end{equation}%
These polynomials may be expressed as ($\cite{Mou}):$

\begin{equation}
\phi _{k,m}\left( z\right) =\frac{\left( -1\right) ^{m\wedge k}}{\sqrt{m!k!}}%
\left( m\wedge k\right) !\left\vert z\right\vert ^{\left\vert m-k\right\vert
}e^{-i\left( m-k\right) \theta }L_{m\wedge k}^{\left\vert m-k\right\vert
}\left( z\overline{z}\right) ,\ z=re^{i\theta },  \tag{3.3}
\end{equation}%
where $L_{m}^{\left( \alpha \right) }$ is the generalized Laguerre
polynomial defined by 
\begin{equation}
L_{m}^{\left( \alpha \right) }\left( x\right) =\frac{1}{m!}%
\sum\limits_{k=0}^{m}\frac{1}{k!}\left( -m\right) _{k}\left( k+\alpha
+1\right) _{m-k}x^{k}.  \tag{3.4}
\end{equation}%
The kernel of the transformation $T_{\tau ,m}$ is therefore given by%
\begin{equation}
W_{\tau ,n}\left( z,w\right) =\sum\limits_{k=0}^{\infty }h_{\tau ,k}\left(
w\right) \overline{\phi _{k,n}\left( z\right) }  \tag{3.5}
\end{equation}%
\begin{equation}
=\sum\limits_{k=0}^{+\infty }\frac{\left( -1\right) ^{n\wedge k}}{\sqrt{\pi
n!k!}}\left( n\wedge k\right) !\left\vert z\right\vert ^{\left\vert
n-k\right\vert }e^{i\left( n-k\right) \theta }L_{n\wedge k}^{\left\vert
n-k\right\vert }\left( z\overline{z}\right) h_{\tau ,k}\left( w\right) 
\tag{3.6}
\end{equation}%
\textit{Proof of Proposition 1.3. }We split $\left( 3.6\right) $ into two
parts as follows%
\begin{equation}
W_{\tau ,n}\left( z,w\right) =W^{(<\infty )}\left( z,w\right) +W^{\left(
\infty \right) }\left( z,w\right)  \tag{3.7}
\end{equation}%
where 
\begin{equation}
W^{(<\infty )}=\sum\limits_{k=0}^{n-1}\frac{\left( -1\right) ^{k}\sqrt{k!}}{%
\sqrt{\pi n!}}\left\vert z\right\vert ^{n-k}e^{i\left( n-k\right) \theta
}L_{k}^{n-k}\left( z\overline{z}\right) h_{\tau ,k}\left( w\right)  \tag{3.8}
\end{equation}%
\begin{equation*}
-\sum\limits_{k=0}^{n-1}\frac{\left( -1\right) ^{n}\sqrt{n!}}{\sqrt{\pi k!}}%
\left\vert z\right\vert ^{k-n}e^{i\left( n-k\right) \theta
}L_{n}^{k-m}\left( z\overline{z}\right) h_{\tau ,k}\left( w\right)
\end{equation*}%
By using the identity $\left( \text{\cite{Sz}},\text{p.98}\right) $ :%
\begin{equation}
L_{j}^{\left( -s\right) }\left( t\right) =\frac{\left( j-s\right) !}{j!}%
\left( -t\right) ^{s}L_{j-s}^{\left( s\right) }\left( t\right) ,\text{ \ }%
1\leq s\leq j  \tag{3.9}
\end{equation}%
for $s=k-m$ and $t=z\overline{z}$, we obtain that

\begin{equation}
\left( -1\right) ^{k}\frac{\sqrt{k!}}{\sqrt{m!}}\left\vert z\right\vert
^{m-k}e^{i\left( m-k\right) \theta }\left[ L_{k}^{m-k}\left( z\overline{z}%
\right) \right] =\left( -1\right) ^{m}\frac{\sqrt{m!}}{\sqrt{k!}}e^{i\left(
m-k\right) \theta }\left\vert z\right\vert ^{k-m}L_{m}^{\left( k-m\right)
}\left( z\overline{z}\right) .  \tag{3.10}
\end{equation}%
Consequently, 
\begin{equation}
W^{(<\infty )}=0,  \tag{3.11}
\end{equation}%
and we are left with 
\begin{equation}
W_{\tau ,n}\left( z,w\right) =W^{\left( \infty \right) }\left( z,w\right)
=\left( -1\right) ^{n}\sqrt{n!}\sum\limits_{m=0}^{+\infty }\frac{1}{\sqrt{m!}%
}\left\vert z\right\vert ^{m-n}e^{i\left( n-m\right) \theta
}L_{n}^{m-n}\left( z\overline{z}\right) h_{\tau ,m}\left( w\right) 
\tag{3.12}
\end{equation}%
\begin{equation}
=\left( -1\right) ^{n}\sqrt{n!}\frac{1}{\overline{z}^{n}}\sum%
\limits_{m=0}^{+\infty }\frac{1}{m!}\overline{z}^{m}L_{n}^{m-n}\left( z%
\overline{z}\right) \left( \frac{\tau }{2}\right) ^{m/2}H_{m}\left( \frac{w}{%
\sqrt{2\tau }}\right) .  \tag{3.13}
\end{equation}%
Let 
\begin{equation}
\mathcal{S}^{\left( \infty \right) }\left( z,w\right)
=\sum\limits_{m=0}^{+\infty }\frac{1}{m!}\left( \sqrt{\frac{\tau }{2}}%
\overline{z}\right) ^{m}L_{n}^{m-n}\left( z\overline{z}\right) H_{m}\left( 
\frac{w}{\sqrt{2\tau }}\right)  \tag{3.14}
\end{equation}%
and use the expansion of the generalized Laguerre polynomial $\left(
3.4\right) $ to rewrite $\mathcal{S}^{\left( \infty \right) }\left(
z,w\right) $ as%
\begin{equation}
\mathcal{S}^{\left( \infty \right) }\left( z,w\right) =\frac{1}{n!}%
\sum\limits_{k=0}^{n}\frac{\left( -n\right) _{k}}{k!}\left\vert z\right\vert
^{2k}\sum\limits_{m=0}^{+\infty }\frac{1}{m!}\left( m+k-n+1\right)
_{n-k}\left( \sqrt{\frac{\tau }{2}}\overline{z}\right) ^{m}H_{m}\left( \frac{%
w}{\sqrt{2\tau }}\right) .  \tag{3.15}
\end{equation}%
Now, notice that the shifted factorial $\left( m+k-n+1\right) _{n-k}=0$ for
any $m<n-k$ while 
\begin{equation}
\left( m+k-n+1\right) _{n-k}=\frac{m!}{\left( m+k-n\right) !},\text{ }m\geq
n-k,\text{ }0\leq k\leq n.  \tag{3.16}
\end{equation}%
Consequently,%
\begin{equation}
\mathcal{S}^{\left( \infty \right) }\left( z,w\right) =\frac{1}{n!}%
\sum\limits_{k=0}^{n}\frac{\left( -n\right) _{k}}{k!}\left\vert z\right\vert
^{2k}\left( \sqrt{\frac{\tau }{2}}\overline{z}\right)
^{n-k}\sum\limits_{m=0}^{+\infty }\frac{1}{m!}\left( \sqrt{\frac{\tau }{2}}%
\overline{z}\right) ^{m}H_{m+n-k}\left( \frac{w}{\sqrt{2\tau }}\right) . 
\tag{3.17}
\end{equation}%
The inner series may be expressed as $\left( \cite{Sin-Sri}\text{, Eq.20}%
\right) :$

\begin{equation}
\sum\limits_{m=0}^{+\infty }\frac{1}{m!}\left( \sqrt{\frac{\tau }{2}}%
\overline{z}\right) ^{m}H_{m+n-k}\left( \frac{w}{\sqrt{2\tau }}\right) =e^{%
\overline{zw}-\tau \overline{z}^{2}/2}H_{n-k}\left( \frac{w}{\sqrt{2\tau }}-%
\sqrt{\frac{\tau }{2}}\overline{z}\right) .  \tag{3.18}
\end{equation}%
Finally,%
\begin{equation}
\mathcal{S}^{\left( \infty \right) }(z,w)=\left( \sqrt{\frac{\tau }{2}}%
\overline{z}\right) ^{n}\frac{e^{\overline{zw}-\tau \overline{z}^{2}/2}}{n!}%
\sum\limits_{k=0}^{n}\left( 
\begin{array}{c}
n \\ 
k%
\end{array}%
\right) \left( -\sqrt{\frac{2}{\tau }}z\right) ^{k}H_{n-k}\left( \frac{w}{%
\sqrt{2\tau }}-\sqrt{\frac{\tau }{2}}\overline{z}\right)  \tag{3.19}
\end{equation}%
and appealing again to $\left( 2.44\right) ,$ we end up with

\begin{equation}
\mathcal{S}^{\left( \infty \right) }\left( z,w\right) =\left( \sqrt{\frac{%
\tau }{2}}\overline{z}\right) ^{n}\frac{e^{\overline{zw}-\tau \overline{z}%
^{2}/2}}{n!}H_{n}\left( \frac{w}{\sqrt{2\tau }}-\sqrt{\frac{\tau }{2}}%
\overline{z}-\frac{1}{\sqrt{2\tau }}z\right)  \tag{3.20}
\end{equation}%
Equivalently,%
\begin{equation*}
W_{\tau ,n}\left( z,w\right) =\left( -1\right) ^{n}\sqrt{n!}\frac{1}{%
\overline{z}^{n}}\mathcal{S}^{\left( \infty \right) }\left( z,w\right)
\end{equation*}%
\begin{equation}
=\left( -\sqrt{\frac{\tau }{2}}\right) ^{nn}\frac{e^{\overline{zw}-\tau 
\overline{z}^{2}/2}}{\sqrt{n!}}H_{n}\left( \frac{w-z}{\sqrt{2\tau }}-\sqrt{%
\frac{\tau }{2}}\overline{z}\right)  \tag{3.21}
\end{equation}%
and the proposition is proved.

\subsection{A quantum mechanical interpretation of the kernel $W_{\protect%
\tau ,n}\left( z,w\right) $}

Le us recall the squeeze operator%
\begin{equation}
S\left( \xi \right) =e^{\xi K_{+}-\overline{\xi }K_{-}},\text{ \ \ \ }\xi
\in \mathbb{C}  \tag{4.1}
\end{equation}%
where $K_{+},K_{-}$ are two of the usual generators of the $SU\left(
1,1\right) $ group, which together with the third, $K_{0},$ satisfy the
well-known commutation relations%
\begin{equation}
\left[ K_{-},K_{+}\right] =2K_{0},\text{ \ }\left[ K_{0},K_{\pm }\right]
=\pm K_{\pm }  \tag{4.2}
\end{equation}%
and may be expressed, in the Bargmann-like representation $\cite{Dat}$, as%
\begin{equation}
K_{+}=\frac{1}{2}z^{2},\text{ \ \ }K_{-}=\frac{1}{2}\partial _{z}^{2},\text{
\ }K_{0}=\frac{1}{2}\left( z\partial _{z}+\frac{1}{2}\right)  \tag{4.3}
\end{equation}%
The operator $S\left( \xi \right) $ is unitary. That is, 
\begin{equation}
S\left( \xi \right) ^{\ast }=S\left( \xi \right) ^{-1}=S\left( -\xi \right) ,%
\text{ \ }\xi \in \mathbb{C}  \tag{4.4}
\end{equation}%
For our purpose, we may consider $\xi \geq 0$ and set $\tau =\tanh \xi $.
Next, we recall the distanglement formula 
\begin{equation}
e^{\xi K_{+}-\overline{\xi }K_{-}}=e^{\tau K_{+}}e^{\left( Log\left(
1-\left\vert \tau \right\vert ^{2}\right) \right) K_{0}}e^{-\overline{\tau }%
K_{-}}  \tag{4.5}
\end{equation}%
used by the authors ($\cite{AGHS},$Theorem.5.1) who proved that%
\begin{equation}
S\left( \xi \right) \left[ w\mapsto \frac{w^{k}}{\sqrt{k!}}\right] =\left(
1-\left\vert \tau \right\vert ^{2}\right) ^{\frac{1}{4}}e^{\frac{1}{2}\tau
w^{2}}\left( \frac{\overline{\tau }}{2}\right) ^{^{\frac{1}{2}k}}\frac{1}{%
\sqrt{k!}}H_{k}\left( \sqrt{\frac{1-\left\vert \tau \right\vert ^{2}}{2%
\overline{\tau }}}w\right)  \tag{4.6}
\end{equation}%
By another side, the kernel $W_{\tau ,n}$ expands as 
\begin{equation}
W_{\tau ,n}\left( z,w\right) =\sum\limits_{k=0}^{\infty }\overline{\phi
_{k,n}\left( z\right) }\left( \frac{\tau }{2}\right) ^{\frac{1}{2}%
k}H_{k}\left( \frac{w}{\sqrt{2\tau }}\right)  \tag{4.7}
\end{equation}%
or, equivalently,%
\begin{equation}
\left( 1-\left\vert \tau \right\vert ^{2}\right) ^{\frac{1}{4}}e^{\frac{1}{2}%
\tau w^{2}}W_{\tau ,n}\left( z,\sqrt{1-\tau ^{2}}w\right) =  \tag{4.8}
\end{equation}%
\begin{equation*}
\sum\limits_{k=0}^{\infty }\overline{\phi _{k,n}\left( z\right) }\left(
1-\left\vert \tau \right\vert ^{2}\right) ^{\frac{1}{4}}e^{\frac{1}{2}\tau
w^{2}}\left[ \frac{1}{\sqrt{k!}}\left( \frac{\tau }{2}\right) ^{\frac{1}{2}%
k}H_{k}\left( \frac{\sqrt{1-\tau ^{2}}w}{\sqrt{2\tau }}\right) \right] .
\end{equation*}%
Therefore,%
\begin{equation}
\left( 1-\left\vert \tau \right\vert ^{2}\right) ^{\frac{1}{4}}e^{\frac{1}{2}%
\tau w^{2}}W_{\tau ,n}\left( z,\sqrt{1-\tau ^{2}}w\right)
=\sum\limits_{k=0}^{\infty }\overline{\phi _{k,n}\left( z\right) }\left[
S\left( \xi \right) \left[ w\mapsto \frac{w^{k}}{\sqrt{k!}}\right] \right] 
\tag{4.9}
\end{equation}%
\begin{equation}
=S\left( \xi \right) \left[ w\mapsto \sum\limits_{k=0}^{\infty }\overline{%
\phi _{k,n}\left( z\right) }\frac{w^{k}}{\sqrt{k!}}\right]  \tag{4.10}
\end{equation}%
where the inner sum, viewed as a function of $w$,%
\begin{equation}
w\mapsto \sum\limits_{k=0}^{\infty }\overline{\phi _{k,n}\left( z\right) }%
\frac{w^{k}}{\sqrt{k!}}  \tag{4.11}
\end{equation}%
belongs to $\mathcal{A}_{n}\left( \mathbb{C}\right) $ and also turns out to
be the expansion (with polyanalytic coefficients $\left\{ \phi _{k,n}\left(
z\right) \right\} $ within the Hilbertian probabilistic scheme $\cite{Gaz})$
of a non normalized coherent state. In this respect, one also notices that
the basis vectors $\frac{1}{\sqrt{k!}}w^{k}=B_{0}\left[ h_{k}\right] \left(
w\right) $ are Bargmann transforms $B_{0}:L^{2}\left( \mathbb{R}\right)
\rightarrow \mathcal{A}_{0}\left( \mathbb{C}\right) $ of Hermite functions $%
h_{k}\left( x\right) $ on $\mathbb{R}$. The latter ones are the\textit{\
number states} of the harmonic oscilator. Therefore, we may normalize the
wavefunction in $(4.11)$ as%
\begin{equation}
\langle w\left\vert z,n\right\rangle =e^{-\frac{1}{2}\left\vert z\right\vert
^{2}}\sum\limits_{k=0}^{\infty }\overline{\phi _{k,n}\left( z\right) }\frac{%
w^{k}}{\sqrt{k!}},  \tag{4.12}
\end{equation}%
and rewrite it as 
\begin{equation}
\langle w\left\vert z,n\right\rangle =B_{0}\left[ x\longmapsto e^{-\frac{1}{2%
}\left\vert z\right\vert ^{2}}\sum\limits_{k=0}^{\infty }\overline{\phi
_{k,n}\left( z\right) }h_{k}\left( x\right) \right] \left( w\right) . 
\tag{4.13}
\end{equation}%
Whence, we recognize the normalized wavefunction of the displaced Fock state
with respect to the Schr\"{o}dinger represention $D\left( z\right) $ of the
Heisenberg group $\mathbb{H}_{1}$ in $L^{2}\left( \mathbb{R}\right) $ $%
\left( \cite{Mou2}\right) :$%
\begin{equation}
\langle x\left\vert z,n\right\rangle =e^{-\frac{1}{2}\left\vert z\right\vert
^{2}}\sum\limits_{k=0}^{\infty }\overline{\phi _{k,n}\left( z\right) }%
h_{k}\left( x\right) =D\left( z\right) \left[ h_{n}\right] \left( x\right) .
\tag{4.14}
\end{equation}%
Therefore, $\left( 4.13\right) $ takes the form 
\begin{equation}
\langle w\left\vert z,n\right\rangle =B_{0}\left[ x\longmapsto \langle
x\left\vert z,n\right\rangle \right] \left( w\right) =B_{0}\left[ x\mapsto
D\left( z\right) \left[ h_{n}\right] \left( x\right) \right] \left( w\right)
.  \tag{4.15}
\end{equation}%
Combining all these facts together, we arrive at 
\begin{equation}
\widetilde{W}_{\tau ,n}\left( z,w\right) :=e^{-\frac{1}{2}\left\vert
z\right\vert ^{2}}\left( 1-\left\vert \tau \right\vert ^{2}\right) ^{\frac{1%
}{4}}e^{\frac{1}{2}\tau w^{2}}W_{\tau ,n}\left( z,\sqrt{1-\tau ^{2}}w\right)
=S\left( \xi \right) \left[ w\mapsto \langle w\left\vert z,n\right\rangle %
\right]  \tag{4.16}
\end{equation}%
which means that this kernel is the squeezed Bargmann analytic
representation of\ displaced Fock states $D\left( z\right) \left[ h_{n}%
\right] $ or, in other words,

\begin{equation}
\widetilde{W}_{\tau ,n}\left( z,w\right) =S\left( \xi \right) \left[
w\mapsto B_{0}\left[ x\mapsto D\left( z\right) \left[ h_{n}\right] \left(
x\right) \right] \left( w\right) \right] .  \tag{4.17}
\end{equation}%
Note that the lowest Landau level $n=0,$ the kernel $\left( 4.17\right) $
reduces further to 
\begin{equation}
\widetilde{W}_{\tau ,0}\left( z,w\right) =\left( 1-\tau ^{2}\right) ^{\frac{1%
}{4}}\exp \left( \frac{1}{2}\tau w^{2}+\overline{z}w-\frac{1}{2}\tau 
\overline{z}^{2}\right) e^{-\frac{1}{2}\left\vert z\right\vert ^{2}}. 
\tag{4.18}
\end{equation}%
But since $w\mapsto $ $\widetilde{W}_{\tau ,0}\left( z,w\right) $ belongs to 
$\mathcal{A}_{0}\left( \mathbb{C}\right) \subset $ $L^{2}\left( \mathbb{C}%
,e^{-\left\vert w\right\vert ^{2}}dw\right) $, we may transfer it back into $%
L^{2}\left( \mathbb{C},dw\right) $ via the inverse ground state
transformation $f\mapsto e^{-\frac{1}{2}\left\vert w\right\vert ^{2}}f$ \ as
follows 
\begin{equation}
e^{-\frac{1}{2}\left\vert w\right\vert ^{2}}\widetilde{W}_{\tau ,0}\left(
z,w\right) =\left( 1-\tau ^{2}\right) ^{\frac{1}{4}}\exp \left( \frac{1}{2}%
\tau w^{2}+\sqrt{1-\tau ^{2}}\overline{z}w-\frac{1}{2}\tau \overline{z}%
^{2}\right) e^{-\frac{1}{2}\left\vert z\right\vert ^{2}-\frac{1}{2}%
\left\vert w\right\vert ^{2}}.  \tag{4.19}
\end{equation}%
Next, by introducing the parameters $\left( a,b\right) $ such that 
\begin{equation}
\tau =\tanh \xi =\frac{\sinh \xi }{\cosh \xi }=\frac{b}{a},\text{ }\left(
1-\tau ^{2}\right) =\frac{1}{a^{2}}  \tag{4.20}
\end{equation}%
Then $\left( 5.35\right) $ may be rewritten%
\begin{equation}
e^{-\frac{1}{2}\left\vert w\right\vert ^{2}}\widetilde{W}_{\tau ,0}\left(
z,w\right) =\frac{1}{\sqrt{a}}\exp \left( \frac{1}{2a}\left( 2\overline{z}w-b%
\overline{z}^{2}+bw^{2}\right) \right) e^{-\frac{1}{2}\left\vert
z\right\vert ^{2}-\frac{1}{2}\left\vert w\right\vert ^{2}}\equiv \langle
z\left\vert w\right\rangle _{g}.  \tag{4.21}
\end{equation}%
This last expression was obtained in (\cite{YuE}, Eq.3.20) as the
wavefunction of the\textit{\ two-photon coherent state} (TPCS)\textbf{\ }of
the radiation field. There, the concept of TPCS was introduced in order to
generalize the Glauber \cite{Glau}\ coherent state 
\begin{equation}
A\left\vert \zeta \right\rangle =\zeta \left\vert \zeta \right\rangle 
\tag{4.22}
\end{equation}%
by replacing the anhilation operator $A$ by its Bogoliubov transform $\left( 
\cite{BV}\right) :$ 
\begin{equation}
A_{g}=aA+bA^{\ast },\text{ }a^{2}-b^{2}=1  \tag{4.23}
\end{equation}%
and the generalized coherent state $\left\vert w\right\rangle _{g}$ was
defined as eigenfunction of the new annihilation operator $A_{g}$ as 
\begin{equation}
A\left\vert w\right\rangle =w\left\vert w\right\rangle _{g}.  \tag{4.24}
\end{equation}%
The wavefunction $\langle z\left\vert w\right\rangle _{g}$ in $\left(
4.21\right) $ has been expressed in the coherence $z$-coordinate. In other
words, $\langle z\left\vert w\right\rangle _{g}$ is the \textit{canonical
coherent state representation }of the TPCS, given by the scalar product $%
\langle z\left\vert w\right\rangle _{g}$ of the TPCS\ $\left\vert
w\right\rangle _{g}$ and the classical Glauber coherent state $\left\vert
z\right\rangle $ of the harmonic oscillator $A^{\ast }A,$ which solves the
problem $A\left\vert z\right\rangle =z\left\vert z\right\rangle .$

Note also that the kernel $\left( 4.21\right) $ corresponds to the \textit{%
Metaplectic representation} of the group $SU\left( 1,1\right) $ in the
Bargmann-Fock space $\mathcal{A}_{0}\left( \mathbb{C}\right) ,$see (\cite%
{Fol}, p.181) and \cite{ITZ}.

\bigskip 

\end{document}